# A system for coarse-grained location-based synchronisation

André Coelho[1], Hugo Ribeiro[1], Mário Silva[1], Rui José[2]

[1]Mestrado em Informática, Universidade do Minho
[2]Departamento de Sistemas de Informação, Universidade do Minho
andresilvacoelho@sapo.pt, hugomsribeiro84@gmail.com, mario_jsilva@netcabo.pt, rui@dsi.uminho.pt

**Abstract.** This paper describes a system for supporting coarse-grained location-based synchronisation. This type of synchronisation may occur when people need only some awareness about the location of others within the specific context of an on-going activity. We have identified a number of reference scenarios for this type of synchronisation and we have implemented and deployed a prototype to evaluate the type of support provided. The results of the evaluation suggest a good acceptance of the overall concept, indicating that this might be a valuable approach for many of the indicated scenarios, possibly replacing or complementing existing synchronisation practices.

## 1 Introduction

Daily life is full of situations in which we have to synchronise our actions with other people. This is an integral part of social interaction and may occur in the context of very diverse social situations. For example, parents need to coordinate to get their kids from school, work colleagues may want to go to lunch together and friends may want to meet at their favourite place. Calendars and agendas are the primary tool for synchronising with others as they enable us to plan and anticipate synchronisation. However, they only represent the expectation that events will happen in a particular way at a particular moment. More recently, mobile phones have also become an important synchronisation tool, allowing people to make only basic arrangements, like "We will meet tonight at one of those bars", and then fine-tune the synchronisation process through the situated exchange of phone calls or SMSs. In fact, many phone calls start by some variant of the question "where are you", especially when there is some expectation that the other person might be somewhere nearby. These forms of synchronisation may reflect pragmatic needs associated with people finding each other, but they are also often just a reflection of a need for reassurance and connectedness towards other people [1].

In this work, we explore how the increasing ubiquity of mobile technologies may support new forms of synchronisation [2]. In particular, we explore the concept of synchronised activity as some type of social activity in which multiple people are involved and desire to maintain a coarse-grained location-based synchronisation between each other. A synchronised activity associates calendar data with a certain physical scope and a set of participants. It provides the context in which those participants will be able to generate and receive relevant location-based notifications that will allow them to perceive how the activity is unfolding in terms of the location of the other participants. We hypothesise that this may extend the role of calendars from planning tools to situated synchronisation tools, and thus from a focus on plans to a focus on situated action as the essence of interpersonal synchronisation [3].

In this paper, we describe the study we have conducted to explore the viability of this concept and gain a more in-depth understanding about its potential and limitations. We start by reviewing systems that share similar objectives and concepts. We then analyse some of the main reference scenarios that we have identified for this type of synchronisation. These scenarios provided the basis for the identification of the requirements of our proposed systems for enabling synchronised activities, which we describe in Sec. 3. Based on these requirements, we propose an architecture to support these synchronisation models and we have created a prototype implementation of that system, both described in Sec. 4. In Sec. 5, we describe the evaluation procedures in which six users have tried the system simulating realistic social situations. The results highlight some important findings, but overall they seem to confirm the validity and opportunity of the concept of synchronised activity.

## 2   Related Work

Our work has some similarities with location sharing systems, like Locaccino or Google Latitude. Locaccino [4] is an application for desktop computers and mobile devices that enables users to share their location with people from their Facebook social network. Location sensing can be done through the use of Wi-Fi or GPS. This system puts great emphasis on the user's privacy. People can set themselves as undetectable whenever desired and they can express the location disclosure preferences using multiple variables, such as time spans and the locations where a group of people is allowed to know the user's location.

Google Latitude [5] is a location-aware application for both desktop and mobile devices that enables users to share their location with their friends with Google accounts. On the desktop, location sensing is achieved through IP geolocation while on mobile devices it is achieved through cellular positioning and GPS. This system, much like Locaccino, puts great emphasis on privacy, reason for which it offers varying degrees of precision in location information, according to what users have chosen to show to other users. Location information can be as precise as the exact GPS coordinates location or as vague as just the city name. The system also overrides old location data with new one so as to avoid the possibility of a user's activity being tracked, unless the user specifically tells the system to keep a history of his locations. The system allows for users to contact users with whom they share their location via Google Talk. This facilitates the possibility of users synchronising for some activity, especially if they are near each other.

The location sharing features of these systems also provide the ground for multiple forms of location-based synchronisation. However, in these systems, synchronisation comes as a by-product of the system's features and not as an integrated part of the tool's design. As a consequence, many of our targets scenarios cannot be properly supported or can only be supported with strongly negative consequence in terms of privacy. In our work, we do not intend to make users traceable all the time, our system is only meant to alert other users of the system when someone has arrived at a predetermined location in the context of some prearranged activity. The fact that location data is only used for the purpose of synchronising people within the scope of a specific activity, together with the potential anonymity of many of our scenarios, means that privacy is much easier to handle in our system than it is within any general purpose location sharing system.

From the perspective of supporting structured awareness about the activity of others, our system also has similarities with several types of ambient display systems. The Whereabouts Clock [6] is a system composed of an ambient display tied to a computer/SMS gateway and a mobile application. The ambient display works as a situated awareness device enabling onlookers to have a persistent, dynamic and at-a-glance view of other peoples' whereabouts. For this purpose the researchers used the clock design metaphor, divided into three portions, each indicating a user's presence in a different location, "in the building", "at home" and "out". Location sensing is achieved through the identification of GSM cells in the user's current vicinity and the different locations indicated by the display must be registered once in the mobile application. In addition users can also broadcast their activity, choosing from a specific list. This system enabled it's users to feel imbued with a sense of remote presence awareness and connectedness.

The key difference to our work is that the whereabouts clock is designed to stay in the same location, the home, and to inform the people that are in that space. Our work is very different in this respect, its purpose is to make such information available to a user anytime and anywhere, in essence, empowering users by making the information mobile. In spite of these differences, the study behind the Whereabouts Clock still allowed us to extract valuable insight, namely the need for the users of the system to understand the context of the information they receive about others and to be able adapt the system to suit their needs.

HomeNote [7] is a system that consists of a software application installed on a tablet PC which is used as a situated display in the houses of families chosen to test the application. The display can receive SMS messages and users can write notes by hand using the tablet PC's stylus. HomeNote aims to exploit the potential and value of person-to-place communication, as opposed to person-to-person communication, in a family environment. With this they aimed to extend their comprehension on the types of communication interactions that are carried on in a family environment and develop support for remote and local situated messaging. The system was regularly used for purposes of synchronisation in the context of an activity. Over the course of the study, tests showed that there were seven types of messages that were common amongst all the households where the system was tested. From those messages we would like to call attention to the following ones: Call for Action, Awareness and Reassurance, Social Touch and Reminders. These are the types of messages whose content and social implications replicate the type of human interaction that our system intends to support and that are most relevant in the scope of synchronisation between people.

The authors conclude that by paying attention to the considerations of some mundane household technologies it is possible to support existing practices and also to create new forms of communication. This is an objective our works share with HomeNote, but on a different perspective. While HomeNote aims to explore people-to-place messaging, our system calls for a more persons-to-persons background interaction. Analysis of this project leads us to believe that, when deploying our application, there is a need to collect information about the extension, quality and diversity of the types of interactions that our application enables.

## 3  Reference Scenarios

In this section, we present a set of reference scenarios that demonstrate possible uses of our system and which have also been used as a basis for requirement identification.

**"Let us meet here in roughly one hour".** This is the scenario where a group of people arrives somewhere and then separates for some time, while doing separate activities. For example, a family may arrive together at a shopping centre. While one of the family members goes to the supermarket, the others will be visiting some local shops. They intend to meet at the end, although they do not know exactly who is going to take longer. Another example may be a tourist bus dropping tourists at a museum. The passengers are expected to be back to the bus after finishing their visit, but the duration of the visit is variable. In this scenario the synchronisation activity is one that is truly very common in everyday life. A group of people separates and agrees to meet at roughly the same time in a designated spot knowing that the subsequent activity is bound by the arrival of all the elements.

**"Who is already there, who is arriving".** In this scenario a store or company organizes a flash mob at some location. They intend to gather a certain number of people in that location, for that effect they might offer some sort of reward for showing up. People adhering to the activity are interested in knowing how many people have shown up already. Another example for this reference scenario is a dinner party. A group of people wanting to get together for dinner, possibly at a restaurant are interested in knowing who has and who has not yet arrived. In such a scenario, synchronisation happens for the effect of gathering multiple people around an activity at a designated location. Synchronisation information here has the role of informing people about activity attendance however, depending on the social context of the activity, the content of such information could come in different forms, due to privacy issues.

**"Your ride is arriving".** Two co-workers go to work together in the same car. One of the co-workers offers to pick the other one up at his house, at a specific time.

The person being picked up finds it useful to know whether or not his colleague is close to the pick-up point, so he is better able to time his arrival and avoid spending unnecessary time waiting on the street. Synchronisation in this scenario happens for the purpose of sharing a resource. Information needed for the purpose of synchronisation is more vital to one of the interested parties involved than to the other because one depends on the actions of the other in order to achieve his goal.

**"Yes, he already took care of that".** This scenario is typical among family members. The heads of the household always need to be in synch to coordinate their efforts with numerous tasks, picking up the children from school, picking up the laundry from the dry cleaner, grocery shopping, etc. As such there is a need to know how things are and who has done what. Synchronisation in this scenario occurs around an activity that benefits more than one person, but can be carried out by a single individual. Synchronisation comes into play because of the fact that other individuals interested in the outcome of the activity feel interest in getting feedback relative to the activity's status, in order to be reassured that things are going along as planned. For instance if one of the parents picks up the children from school, the other parent will feel a need to know when that happens and if everything goes along well.

## 4   System support for coarse-grained location systems

In this section, we describe the platform that we created to support coarse-grained location-based synchronisation.

### 4.1 Requirements

From the analysis of the previous scenarios we were able to identify the following list of requirements:
- Activity support is bounded by a temporal context in which it is to happen.
- Activity support is tied to the existence of a geographic scope associated with each activity.
- Activities must support the involvement of multiple people.
- The system must enable users to activate/deactivate synchronisation functionalities regarding an activity at a time of their choosing.
- The system must act as mediator because people might not know each other and they might not know of each others' whereabouts, but they must still be able to synchronise in the context of an activity.

Other requirements are tied to details such as configuration parameters and privacy. Depending on the social context of the activity, users may desire to enforce different privacy policies regarding identity disclosure.

### 4.2 Architecture

The architecture we envisioned for our system, represented in Fig. 1, is composed of three distinct entities: a mobile application running on a smartphone with internet connectivity and GPS, a server that handles all notifications to and from users of the mobile application and a shared calendar system, where activities can be specified using common mechanisms to create events in the calendar.

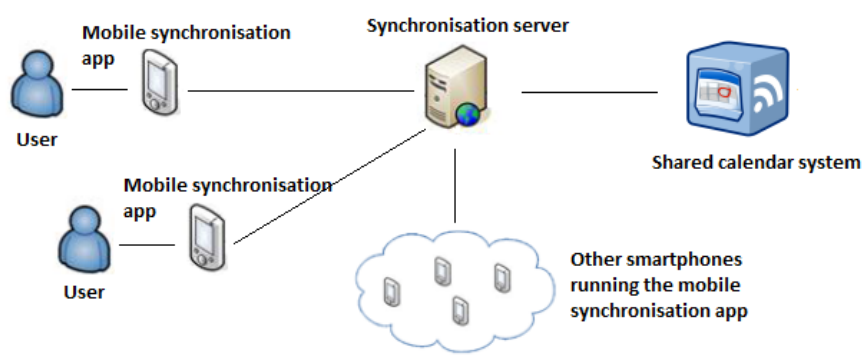

**Figure 1.** System architecture.

#### 4.2.1 Mobile synchronisation application

The mobile synchronisation application is the primary point of entry into the system allowing users to create and manage activities while on the move. The mobile application supports several activity creation models. Users can create an activity while: being physically present at the site; not being present at the site but knowing the location's coordinates beforehand; or, creating an activity without location coordinates and adding them at a later time. Regarding activities the application supports different privacy policies chosen by the user for each activity; these policies will affect issues like identity disclosure. The mobile application is also responsible for warning the server when a user enters the physical region that was associated to an on-going activity. This will cause the server to generate notifications for all participants relevant to the activity and these notifications will be delivered to them via the application in their mobile devices, which effectively makes the application the endpoint for server notifications.

#### 4.2.2 Shared calendar system

The shared calendar system (e.g. Google Calendar) provides an alternative entry point for calendar functionality and participant invitation, enabling people to create activities using a familiar interface. The participants invited to the calendar event will also become the participants in the synchronised activity The only difference for common calendar events is the possibility to encode coordinates in the calendar event and the need to include the system's own e-mail address in the invited list to make the system aware of this new activity.

### 4.2.3 Synchronisation server

The server is the part of the system responsible for receiving user notifications related to activities and generating and forwarding the appropriate notifications to other users in the context of said activities and according to the privacy policies appropriate for each activity type. When invited to a calendar event, the synchronisation server will interface with the shared calendar system and download activity data, such as participants list, location, start time, end time and type of activity. It will then manage the necessary notifications to the mobile synchronisation applications.

### 4.2.4 System operation

When someone creates a new activity, from either the mobile application or the shared calendar system, the indicated participants will be notified through their mobile applications and they can either accept or deny participation in the activity. This allows the synchronisation server to keep track of which guests have accepted or not to participate in an activity, and it allows it to manage activities for the purpose of issuing notifications to users.

When an activity's start time has been reached, the server will start accepting communication from mobile devices regarding that activity. As users enter the geographic region defined for the activity, they will be notified of this occurrence on their devices and the devices will notify the synchronisation server of user arrival. When this happens the synchronisation server will issue notifications to the mobile devices of participants informing them that a user has arrived. Depending on the privacy policy the notification can feature the identity of the person that has arrived.

Different activity types cause different notifications to be generated and with different frequency, for instance, if the activity is a flash mob, the server will notify users of arrivals at a frequency of X arrivals, so as to not annoy users with too many notifications.

## 4.3 Implementation

To support the evaluation of the key concepts proposed in this work, we have implemented a simplified version of our coarse-grained location-based personal synchronisation system. Based on the technical requirements we had for the mobile application, we have implemented that part of the system using the Android platform. This choice is tied to several factors, the main one being that the LBS (location-based services) API seemed very strong and provides an easy way to obtain the desired behaviour for our application in what regards user presence detection in a specified geographical region. The fact that the Android OS allows us to run applications in the background and its power management features, namely the fact that applications and GPS still work while the phone is on standby were also critical to this choice. Other factors that drew us to this platform were: the familiarity with Java and the general tidy, regulated and balanced feel of the programming model as a whole, which we feel results in applications being more suited to a mobile environment's requirements.

This implementation works as a standalone application that is designed to be tested by one individual carrying an Android device. All behaviour related to other users, in the context of an activity, is simulated by the prototype via notifications pertaining to alterations in the system caused by a user's actions, like arriving at the rendezvous point.

## 5   Evaluation

The overall objective of our evaluation was to gain some insight into the viability of the concept of coarse-grained synchronisation and inferring its potential as a method for replacing or complementing existing synchronisation practices. Within this broader objective, we also intended to assess more specific characteristics of our implementation, such as:

- Determine user's ease of use and learning of the application's interface;
- Determine whether the users were able to ascertain the general state of the activities he is involved in, as well as the repercussions of his actions for the system and for the other users of the system, from the perspective of what it means to be in synch with other people in the context of an activity;
- Determine if the system's feedback was appropriate and useful to users in regards of achieving the objective of synchronising with other people.

### 5.1   Methodology

The methodology for this evaluation is composed of a field test followed by a questionnaire. Field testing the application with volunteers was conducted using the following scenario: "*Two family members who are out together decide to split up because one of them wants to embark on an activity that the other is not very keen on undertaking, in this case, shopping at a fair. As a result, they schedule a time and a place to meet up so they can both go their separate ways and use their time as they see fit.*"

The tests were carried in three distinct locations, University of Minho's Campus, Mire de Tibães and Maximinos. All of these locations constitute a viable setting for the occurrence of a synchronisation activity as defined in the scope of our work.

We recruited 6 volunteers, chosen amongst friends whom we felt would be able to deliver a straightforward opinion in their evaluation. They had varying degrees of expertise in interfacing with touch based devices and more specifically the Android OS, which undeniably reflects in their opinions on the usability portion of the test.

After having set the scenario and contextualizing the experiment for them, they were presented with the Android device so they would be able to evaluate the system and draw their own conclusions. Upon finishing the field test and assimilating the experience, users were presented with a questionnaire. The part of the questionnaire related to usability was conducted in the moulds of a publicly available online heuristics questionnaire [8], this was fused together with an additional section of questions pertaining to our other more fundamental evaluation objectives, questions which we thought were relevant and capable of directing the users to providing us the feedback we wanted. Some of the heuristics questions were adapted to better suit the purposes of our evaluation. Each questionnaire was answered in approximately 10 minutes and all information resulting from them stored in digital format. The only audiovisual record created was a demonstration of the working prototype.

### 5.2 Platform and test prototype

The platform used for testing is an HTC mobile device, model name Legend. The device comes equipped with an integrated GPS module, a 3.2" screen with a resolution of 320x480 (HVGA), running Android 2.1 (Éclair) with HTC Sense UI and all the input is touch driven.

The test prototype application uses GPS and the Android LBS API to sense a user's presence in the area chosen for an activity. It provides a help menu so that the user may clarify any doubts regarding what each of the controls do and regarding the process of activating a presence alarm for an activity. All activity data is collected using the calendar component that's provided in the Android OS, which is called from within our own application. This data is then stored in the devices SQLite database.

The following images present screenshots of the test prototype used in evaluation. In Fig. 2a we can see a detailed view of an on-going activity and its status. Fig. 2b shows the user receiving an update, Fig. 2c shows the content of the update, indicating that Mario has arrived to the activity location.

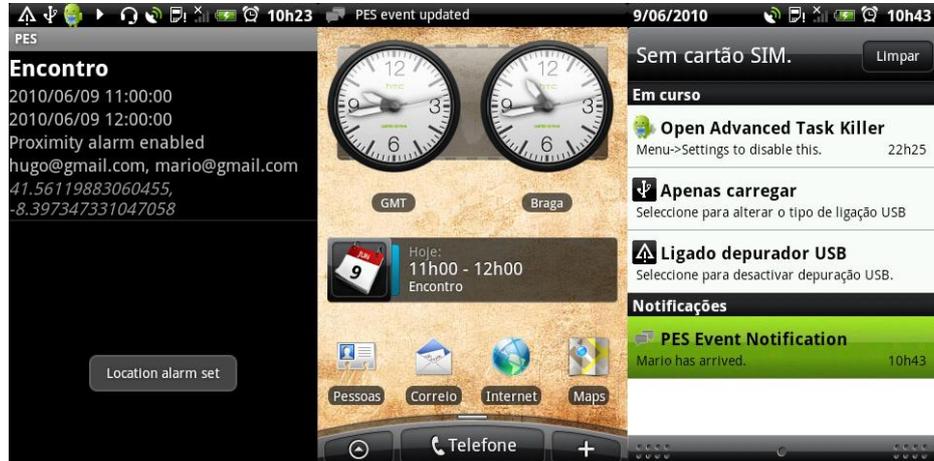

**Figure 2.** (**a**) Detailed view of an activity. (**b**) Statusbar notification for an activity. (**c**) Notifications pane with detailed description of the notification.

### 5.3 Analysis

Overall, the results obtained during the evaluation suggest that the system was positively perceived by the volunteers. In this section, we describe some of the main findings.

#### 5.3.1 Usability

Regarding usability, opinions amongst testers were varied, some felt that the interface was simple, intuitive and to the point, while others felt it needed some refinement and glare. We have perceived that some of the problems that the users have identified were clearly connected with their lack of experience with touch based devices and especially the Android OS's UI interaction model.

Another issued raised by users was directly connected with the use of the Android notification model. We used this notification framework to warn users when someone arrives at the location of the activity. Users expressed concerns over the possibility of missing notifications, due to the default notification sound being too short and also, because there was no vibration or flashing LED warnings.

Other issues that were pointed out by some users regard the approach used to fill out the activity location field with GPS coordinates and the activation of a location alarm. Those users felt there should be a better way to get and set the GPS coordinates of an activity. Initially we had made that process automatic and transparent to the user, which on the face of it might seem ideal, but it posed a serious limitation in the way that it was done because it forced the user to be physically present at the point of rendezvous when setting the alarm, which is not ideal. So we opted for another approach which was to have the users press the GPS Coordinates button, which copies them to the clipboard and then have users paste them in the activity location field. This decoupled coordinate setting and alarm activation, so the users can have more freedom when it comes to the process of setting the meeting point for an activity. Still, we consider none of these approaches to be ideal and the application needs further refinement in this aspect. At the beginning of an activity, when still at the would-be meeting point, the alarm could be immediately activated. In doing so, the system will notify other users of an arrival, which was not the intention.

#### 5.3.2 Main learnings from the study

The overall results suggest that users see great potential in a synchronisation tool, such as this one, as a method for replacing, or in some cases complementing, the traditional forms of synchronisation (SMS/phone call). The data suggests users testing the system find that the feedback given to them by the system is adequate and can easily substitute the one obtained via the traditional methods referred, thus validating our goal of facilitating interpersonal synchronisation in the context of an activity by extending the calendar as a tool of coarse-grained location based synchronisation.

User feedback showed that users are able to accompany and realize what the state of an activity is, as well as the repercussions of their actions, towards others and the system. This tells us that users trust the system and the information it relays to them, which is critical towards application viability in this context.

In spite of these promising results, users still expressed concerns that phones with the necessary capabilities may not be adequately priced, while others consider that the need for internet connectivity may result in them spending more money than they would with an SMS or a phone call. Other users pointed out other potential issues like internet connectivity and GPS connectivity driving down the device's autonomy. This issue is in part addressed by the Android platform itself with its advanced power management features, but it will continue to be mitigated by hardware evolution and also continued software evolution as Android is constantly evolving.

## 6   Conclusions

One of the main objectives was to explore the concept of coarse-grained interpersonal synchronisation using a calendar as an underlying tool and extending the calendar's functionality to provide coarse-grained location based synchronization. This is a goal we believe to have hit with a good measure of success, since the results of our evaluation with users suggest that the application developed to explore this concept seems to hold great potential and value for users.

In our study we were able to identify some issues in the ecosystem that could impact adoption of such a system. Factors like the current market and economic status quo in what regards mobile devices, namely smartphone pricing, adoption rate and internet data plans are a source of concern for users and may affect system adoption. Another key issue pertaining to the ecosystem is related to the architecture of the system, specifically, the shared calendar component. There is a definite need for one, but not obvious solution. Google Calendar is a viable option given that it's widely used and seems like a good approach to the issue, but its API is still not very matured. Ideally, our system should easily integrate with multiple types of calendar system, as we only make a very simple use of their features.

Privacy is also an important issue in such systems but in our research we were not able to determine any critical issues and users presented no objections.

As with any other project there is always work to be done in the future. At this point in time, we can point out an obvious issue to be addressed, which is to implement the remainder of the system. Additionally, there is room for optimizing the way polling to the synchronisation server is done. Factoring in information we already possess about planned activities, like the starting time and the type of activity, one can adjust the frequency with which the application polls the server. This would result in a better usage of battery and data, and would also contribute to the user having more up-to-date data at relevant times. Further into the future, one could extend the application functionalities. For instance, one feature we can see as being useful in a tool like this is to have the application interface with Google Maps to give the person directions to the meeting point. Users pointed out other interesting features like the production of graphics with user attendance and assiduity and possibly the ability to share them with friends.

Overall, we feel our research project is of valuable use to someone wanting to explore the underlying concept of synchronisation and that such an exploration could be carried out using the foundations we have laid down with our work.